\documentclass{article}
\usepackage{spconf,amsmath,graphicx}

\usepackage{color}
\usepackage{enumitem}
\usepackage{hyperref}
\usepackage{url}
\usepackage{breakurl}
\usepackage{multirow,makecell}
\usepackage{etoolbox,siunitx}
\robustify\bfseries
\usepackage{booktabs}
\usepackage{balance}
\usepackage{amssymb}
\usepackage{bm}
\usepackage{arydshln}
\usepackage{caption}
\usepackage{nicefrac}
\usepackage{cite}
\usepackage{cite}
\usepackage{bbm}
\usepackage[scr=boondoxo]{mathalfa}

\usepackage{pifont}% http://ctan.org/pkg/pifont
\let\oldding\ding% Store old \ding in \oldding
\renewcommand{\ding}[2][1]{\scalebox{#1}{\oldding{#2}}}% Scale \oldding via optional argument

\def\s{\mathbf{s}}

\usepackage{amssymb}% http://ctan.org/pkg/amssymb
\usepackage{pifont}% http://ctan.org/pkg/pifont
\newcommand{\xmark}{\ding{55}}%

\title{Latent Iterative Refinement for Modular Source Separation}

\name{Dimitrios Bralios$^1$\thanks{D. Bralios and P.\ Smaragdis were partially funded by NIFA grant \#2020-67021-32799. E.\ Tzinis was partially funded by the Google Ph.D.\ fellowship.
Code: \href{https://github.com/dbralios/latent_iterative_refinement_sep}{https://github.com/dbralios/latent\_iterative\_refinement\_sep}}
, Efthymios Tzinis$^1$, Gordon Wichern$^2$,  Paris Smaragdis$^{1}$, Jonathan Le Roux$^2$
\address{$^{1}$University of Illinois Urbana-Champaign, Urbana, IL, USA \\
$^2$Mitsubishi Electric Research Laboratories (MERL), Cambridge, MA, USA}}
\begin{document}
\ninept
\maketitle
\setlength{\abovedisplayskip}{3pt}
\setlength{\belowdisplayskip}{3pt}
\begin{abstract}
%  Traditional source separation approaches train deep neural network models end-to-end with all the data available at once by minimizing the empirical risk on the whole training set. Moreover, on the inference side, after training the model, the user fetches a static computation graph and runs the full model on some specified observed signal to get a reconstruction of the true uncorrupted signal. We argue that we can significantly reduce the computation during both training and inference stages by reformulating model’s training and inference procedures as successive mappings of latent signal representations. During training we can apply the same processing block more than one time on the same input to make a better approximation of the input signal and consequently backpropagate the error through the parameters of the blocks that have been applied. To that end, one can train a very complicated network structure using significantly less computation compared to the end-to-end training and can perform intermediate block augmentation using perturbations of the signal estimates. During inference, we can dynamically adjust how many processing blocks an input-signal needs using a threshold in terms of absolute change in the latent space or some estimate of uncertainty for the current output.
Traditional source separation approaches train deep neural network models end-to-end with all the data available at once by minimizing the empirical risk on the whole training set. On the inference side, after training the model, the user fetches a static computation graph and runs the full model on some specified observed mixture signal to get the estimated source signals. Additionally, many of those models consist of several basic processing blocks which are applied sequentially. We argue that we can significantly increase resource efficiency during both training and inference stages by reformulating a model’s training and inference procedures as iterative mappings of latent signal representations. First, we can apply the same processing block more than once on its output to refine the input signal and consequently improve parameter efficiency. During training, we can follow a block-wise procedure which enables a reduction on memory requirements. Thus, one can train a very complicated network structure using significantly less computation compared to end-to-end training. During inference, we can dynamically adjust how many processing blocks and iterations of a specific block an input signal needs using a gating module.
\end{abstract}

\begin{keywords}
Iterative latent refinement, progressive training, sound source separation, layer-wise optimization
\end{keywords}

\section{Introduction}

Layer-wise optimization is a widely used idea in terms of progressively training a deep neural network using shallow layer-wise losses \cite{bengio2006greedy,karras2017progressive}. This technique has been successfully applied towards applications such as image classification \cite{trinh2019greedy}, image generation \cite{karras2017progressive}, or speech enhancement \cite{gao2018densely,llombart2019progressive, sun2020progressive, kim2022bloom}, and has been shown to perform comparably with end-to-end training but with large computational savings. This is especially important when one can only fit sub-parts of the forward and backward procedure in memory, e.g. a single trainable layer and not the full model.

Recursive neural networks are centered around the idea of weight-sharing in a sequence of layers \cite{socher2011parsing, eigen2013understanding, liang2015recurrent, komatsu2022non}. Therefore, they have been proposed as a way to reduce the overall number of parameters of models. In the case of speech enhancement, \cite{li2020time} proposes a network consisting of multiple stages where feedback from the previous stage is provided. The concept of iterative refinement is also found in more recent diffusion based models  \cite{ho2020denoising, chen2020wavegrad}.

Further, dynamic neural networks and adaptive computation are other broad concepts that have been explored recently \cite{han2021dynamic}. On the inference side, previous works have proposed to train models which are able to perform early-exit and save computation for inputs which are less noisy or sometimes when more denoising could lead to more artifacts that would harm the performance of downstream tasks \cite{kaya2019shallow, gao2018densely, sun2020progressive, chen2021don}. An early exiting mechanism centered around a manually set threshold on a distance between the output of successive processing stages, is proposed for the problem of source separation \cite{chen2021don} as well as for speech enhancement \cite{li2021learning}. In other tasks, different approaches utilize a module that controls the flow of computation allowing data-adaptive computation. For instance, in \cite{Wang2018SkipNet}, a gating module has been proposed in order to controls layer-wise skip connections and is trained using reinforcement learning methods. In \cite{guo2019dynamic}, instead of skipping layers, the gating module controls the number of iterations for some layers, while in \cite{bengio2013estimating} the gating function is trained jointly. 

In this work, we build upon these previous results and follow a more holistic approach towards utilizing both block-wise training (such as Bloom-Net \cite{kim2022bloom}) and inference recipes, with the important addition that we can also perform multiple applications of the same blocks (with the same parameters), % – see Fig.\ \ref{fig:blockiter}
consequently reducing the model size. In the case where we progressively train the network, we enable a much larger parallelization by greatly reducing computational demands during training. 

During the process of refinement, some of the noisy signals approach their ideal targets in a small number of steps even when applying the same layer multiple times. This indicates that a share of the samples can be processed in fewer steps without causing a significant drop in performance. To that end, we can transform the separation model into an adaptive network which determines the amount of computation needed on a sample-by-sample basis. We achieve this by attaching to the separation network a learnable gating module (based on \cite{guo2019dynamic}) that for each input decides when to take an early exit. Joint training of the separation model and the gating module not only renders the network dynamic but it also enables it to adapt to different inputs. Our experiments in both source separation and speech enhancement show the effectiveness of our proposed holistic latent iterative refinement approach.

% Through experiments on both speech enhancement and noisy speech separation, we demonstrate the utility of our proposed approach for resource efficient applications across several dimensions. First, by sharing parameters across computation blocks, we reduce model size. Second, by through progressive training of computational blocks, we reduce the memory requirements for model training. Finally, we reduce computation at inference time by adaptively selecting the number of required computation blocks for each sample (e.g., use more blocks for nosier samples).

\section{Latent Iterative Refinement}
\label{sec:refinement}

\noindent Given an input mixture $\mathbf x$, the objective of a source separation network is to recover the sources $\mathbf s$ that compose it. A large class of source separation architectures fit the following definition:
\begin{equation}
\label{eq:masksep}
    \begin{gathered}
    \hat{\mathbf{m}} = \text{Mask}  \left[ \mathcal{S}  \left( \mathcal{E}(\mathbf {x}) \right)  \right],\\
    \hat{\s} = \mathcal{D} \left( \mathcal{E}\left( \mathbf{x} \right) \odot \hat{\mathbf{m}} \right),
    \end{gathered}
\end{equation}
where $\mathcal{E}$ is an encoder which transforms the time-domain input mixture into a latent representation. This latent representation is then fed into a separation module $\mathcal{S}$ whose output is transformed using a $\text{Mask}$ network into mask estimates $\hat{\textbf{m}}$ which are applied on the mixture's latent representation. Finally, $\mathcal{D}$ is a decoder which turns the latent-space source estimates into the time-domain estimated sources $\hat {\textbf{s}}$. Usually, the separation module $\mathcal{S}$ is a composition of blocks which have the same structure, as shown in Fig.~\ref{fig:sepmodel}:
 \begin{figure}[htb]
    \centering
    \includegraphics[width=\linewidth]{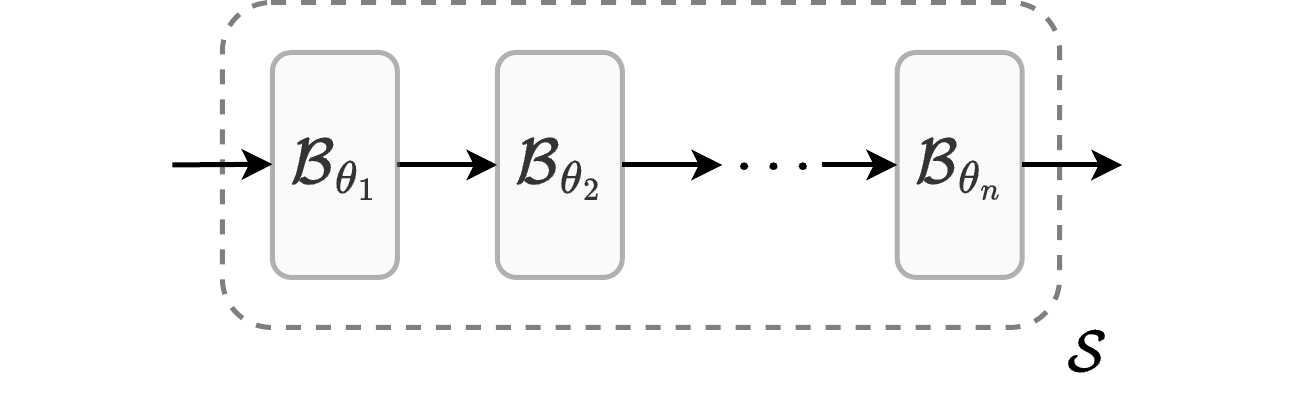}
      \caption{A separation module $\mathcal{S}$ composed of blocks $\mathcal{B}_{\theta_i}$, each of which refines a latent-space estimate.}
      \label{fig:sepmodel}
    % \vspace{-8.8pt}
\end{figure} 
\begin{equation}
\label{eq:sepmodule}
    \begin{gathered}
    \mathcal{S} \left( \mathbf{v} \right) = \left( \mathcal{B}_{\theta_n} \circ \mathcal{B}_{\theta_{n - 1}} \circ \dots \circ \mathcal{B}_{\theta_1}\right) \left( \mathbf{v} \right),
    \end{gathered}
\end{equation}
 where $\circ$ denotes composition, $\mathcal{B}_{\theta_i}$ denotes the $i$-th block with parameters $\theta_i$, and $\mathbf{v} \in \mathbb{R}^{C \times L}$ is the latent mixture representation fed to the separation module. Formulating the separation module this way, we can regard the effect of the blocks as an iterative refinement process of a latent-space estimate. Thus, based on this formulation, we present a unified framework encompassing a series of techniques with multiple uses centered around resource efficiency.

\subsection{Iterative Block-Wise Latent Refinement}
\label{subsec:blockiter}

Casting the separation module as a refinement process, we can impose the constraint of shared parameters between its blocks. In this case, we have $\theta_i := \theta, \,\, \forall i \in \left \{ 1, \dots , n\right \}$, and the separation module becomes as such:
\begin{equation}
\label{eq:sepmoduleiter}
    \begin{gathered}
    \mathcal{S} \left( \mathbf{v} \right) = \big( \underbrace{\mathcal{B}_{\theta} \circ \mathcal{B}_{\theta} \circ \dots \circ \mathcal{B}_{\theta}}_{n \text{ times}}\big) \left( \mathbf{v} \right), % = \mathcal{B}^n_\theta \left( \mathbf{v} \right),
    \end{gathered}
\end{equation}
where in essence the block $\mathcal{B}_\theta$ is applied iteratively to its output. This allows us to drastically reduce the number of model parameters albeit at the cost of a performance drop.
In order to mitigate this reduction in performance, we can increase the size of a block $\mathcal{B}_{\theta_i}$, which can be composed of $k$ distinct sub-blocks $\mathcal{C}_{{\theta}_{i,j}}$ as follows: 
\begin{equation}
\label{eq:subblocks}
    \begin{gathered}
    \mathcal{B}_{\theta_i} \left( \mathbf{v} \right) = \big( {\mathcal{C}_{\theta_{i,k}} \circ \mathcal{C}_{\theta_{i,k-1}} \circ \dots \circ \mathcal{C}_{\theta_{i,1}}}\big) \left( \mathbf{v} \right).
    \end{gathered}
\end{equation}

\begin{figure}[!h]
    \centering
    \includegraphics[trim={1cm 0 1cm 0},clip,width=\linewidth]{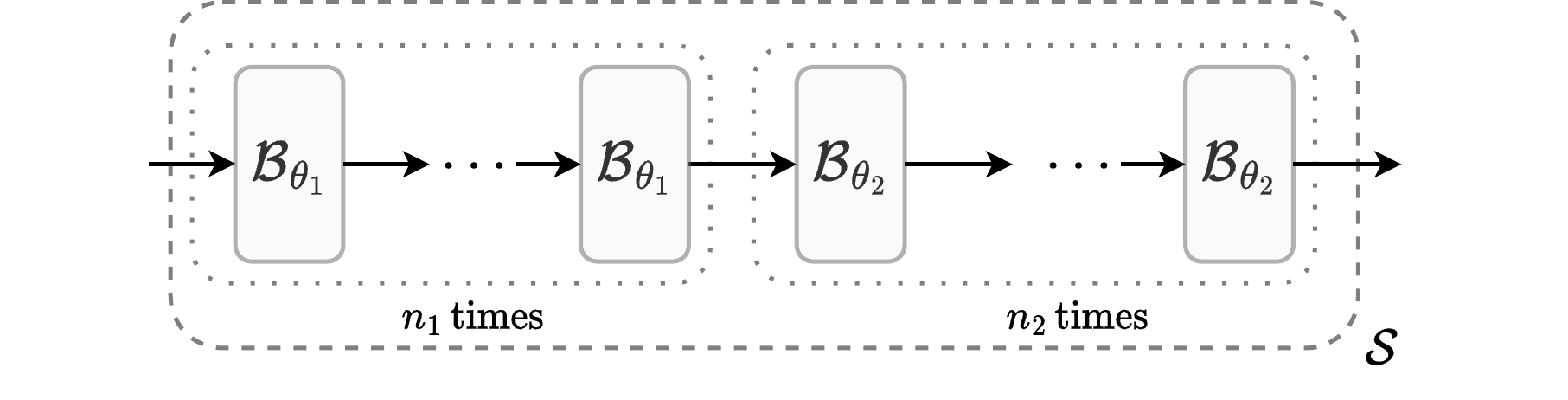}
      \caption{A separation module $\mathcal{S}$ constructed from two different blocks $\mathcal{B}_{\theta_1}$ and $\mathcal{B}_{\theta_2}$ iterated $n_1$ and $n_2$ times respectively. }
      \label{fig:blockiter}
    % \vspace{-8.8pt}
    \vspace{-.2cm}
\end{figure}

Further, we can have more than one iterated block. For example, given the blocks $\mathcal{B}_{\theta_1}$ and $\mathcal{B}_{\theta_2}$, we can construct the following separation module:
\begin{equation}
\label{eq:sepmoduleiter2}
    \begin{gathered}
    \mathcal{S}  \left( \mathbf{v} \right) = \left(\left( \mathcal{B}_{\theta_2} \circ \dots \circ \mathcal{B}_{\theta_2}\right) \circ \left( \mathcal{B}_{\theta_1} \circ \dots \circ  \mathcal{B}_{\theta_1}\right) \right) \left( \mathbf{v} \right),
    \end{gathered}
    % \begin{gathered}
    % \mathcal{S}  \left( \mathbf{v} \right) = \left( \mathcal{B}^{n_2}_{\theta_2}  \circ \mathcal{B}^{n_1}_{\theta_1}  \right) \left( \mathbf{v} \right),
    % \end{gathered}
\end{equation}
which is depicted in Fig.~\ref{fig:blockiter}. In the case of $m$ distinct iterated blocks, the network is defined as:
\begin{equation}
    \begin{gathered}
    \mathcal{S}  \left( \mathbf{v} \right) = \left(\left( \mathcal{B}_{\theta_m} \circ \dots \circ \mathcal{B}_{\theta_m}\right) \circ \dots \circ \left( \mathcal{B}_{\theta_1} \circ \dots \circ  \mathcal{B}_{\theta_1}\right) \right) \left( \mathbf{v} \right).
    \end{gathered}
    % \begin{gathered}
    % \mathcal{S}  \left( \mathbf{v} \right) = \left(\mathcal{B}^{n_m}_{\theta_m}  \circ \dots \circ   \mathcal{B}^{n_1}_{\theta_1} \right) \left( \mathbf{v} \right).
    % \end{gathered}
\label{eq:sepmodule_general}
\end{equation}

\subsection{Progressive Latent Representation Training}
\label{subsec:progtrain}

\begin{figure}[!h]
    \centering
    \includegraphics[trim={0.0cm 0 0.0cm 0},clip,width=\linewidth]{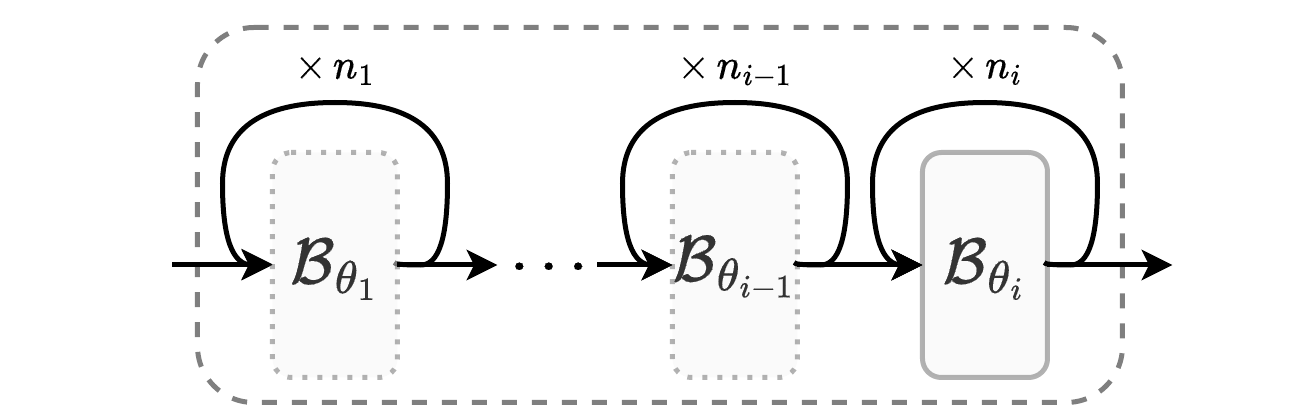}
      \caption{Separation module at step $i$ of the progressive training where block $\mathcal{B}_{\theta_i}$ is trained while the previous iterated blocks $\mathcal{B}_{\theta_1}, \, \mathcal{B}_{\theta_2}, \, \dots , \, \mathcal{B}_{\theta_{i - 1}}$ are kept frozen, as represented by the dotted lines. Feedback connections represent iterations depicted in a more compact way.}
      \label{fig:progtrain}
    \vspace{-.2cm}
\end{figure}  

Instead of learning the complete refinement process in an end-to-end manner, we can split up the training into $m$ different steps, where $m$ is the number of distinct blocks. As before, the separation module $\mathcal{S}$ is given by Eq.~\ref{eq:sepmodule_general} and each block $\mathcal{B}_{\theta_i}$ can be composed of several sub-blocks $\mathcal{C}_{\theta_{i,j}}$ as described in Eq.~\ref{eq:subblocks}.

In the first training step, we learn the shared encoder $\mathcal{E}$ and the first block $\mathcal{B}_{\theta_1}$ along with its dedicated mask network and decoder $\mathcal{D}_1$. After this step is completed, we keep the parameters of the shared encoder $\mathcal{E}$ and the first block $\mathcal{B}_{\theta_1}$ frozen as we proceed. At each of the following steps, a block $\mathcal{B}_{\theta_i}$ along with its dedicated decoder $\mathcal{D}_i$ and its mask network are trained while the previous iterated blocks $\mathcal{B}_{\theta_1}, \, \mathcal{B}_{\theta_2}, \, \dots , \, \mathcal{B}_{\theta_{i - 1}}$ (as shown in Fig.~\ref{fig:progtrain}) and the shared encoder $\mathcal{E}$ are kept fixed.

%$\mathcal{B}^{n_1}_{\theta_1}, \, \mathcal{B}^{n_2}_{\theta_2}, \, \dots , \, \mathcal{B}^{n_{i-1}}_{\theta_{i - 1}}$

This technique allows us to reduce the computational requirements of the training procedure in terms of memory. Additionally, after training is completed, we have $m$ different models of increasing size and computational requirements which can be deployed depending on the available resources.

\subsection{Adaptive Early Exit}
\label{subsec:adaptive}

Instead of having a static refinement process, we can turn it into a dynamic process where the number of processing steps differs for different inputs. We accomplish this by using a gating module $\mathcal{G}_{\phi}$ with parameters ${\phi}$, which, given an input $\mathbf{v}_{i-1} \in \mathbb{R}^{C \times L}$, decides if processing by block $\mathcal{B}_{\theta_i}$ is needed (i.e., $\mathcal{G}_{\phi} \left( \mathbf{v}_{i-1} \right) = 1$) or not (i.e., $\mathcal{G}_{\phi} \left( \mathbf{v}_{i-1} \right) = 0$). Formally, a step of refinement then becomes: 
\begin{multline}
\label{eq:gating}
    F_i\left( \mathbf{v}_{i-1} ; \mathcal{B}_{\theta_i}, \mathcal{G}_{\phi}\right) = \mathcal{B}_{\theta_i} \left( \mathbf{v}_{i-1} \right)  \mathcal{G}_{\phi} \left( \mathbf{v}_{i-1} \right) \\
    + \mathbf{v}_{i-1} \left(1 - \mathcal{G}_{\phi} \left(\mathbf{v}_{i-1} \right)\right).
\end{multline}
Thus, the separation module $\mathcal{S}$ can then be formulated as follows:
\begin{equation}
\label{eq:gating2}
    \begin{gathered}
    \mathcal{S} \left( \mathbf{v} \right) = \left(  F_n \circ  F_{n - 1} \circ \dots \circ  F_1 \right)\left( \mathbf{v} \right).
    \end{gathered}
\end{equation}

\begin{figure}[!h]
    \centering
    \includegraphics[trim={0.5cm 0 0.5cm 0},clip,width=\linewidth]{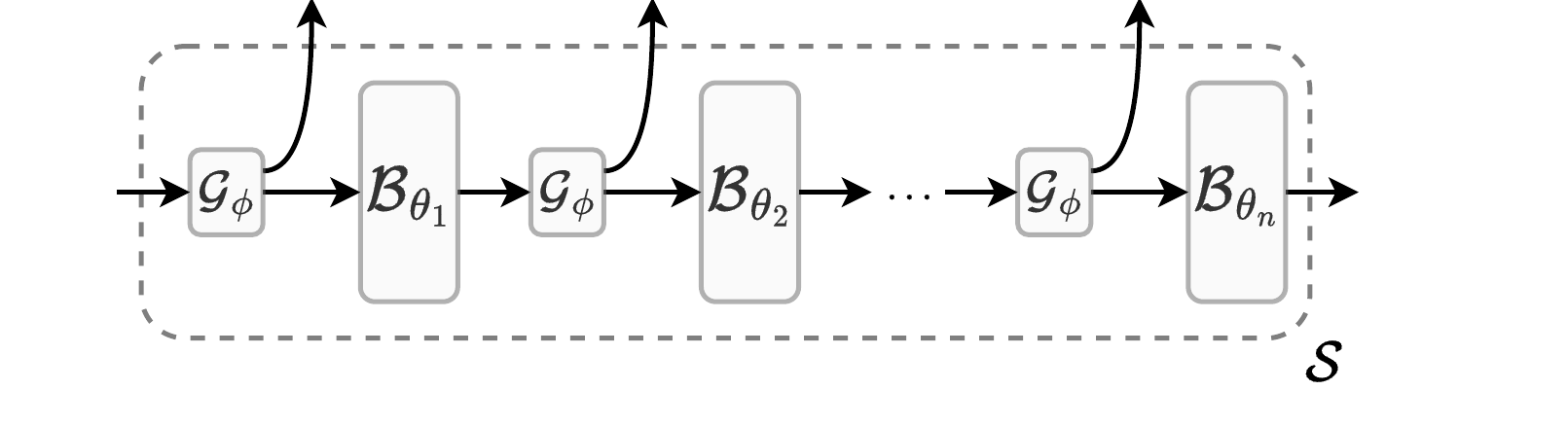}
      \caption{Separation module $\mathcal{S}$ paired with a gating module $\mathcal{G}_{\phi}$ which during inference decides at each step of the refinement process whether an early exit should be taken or not. }
      \label{fig:gating}
    % \vspace{-8.8pt}
\end{figure}

The gating module used in the experiments follows a similar architecture as described in \cite{guo2019dynamic}.
Specifically, 
\begin{equation}
\label{eq:gating3}
    \begin{gathered}
    \mathcal{G}_{\phi} \left( \mathbf{v}_{i-1} \right) = 
    \begin{cases} 
    \text{arg}\,\text{max}\left( f_{\phi} \left(\mathbf{v}_{i-1}\right) + G\right) & \text{Forward pass} \\ 
    \text{softmax}\left( f_{\phi} \left(\mathbf{v}_{i-1}\right) + G \right) & \text{Backward pass}
    \end{cases},
    \end{gathered}
\end{equation}
where $f$ turns $\mathbf{v}_{i-1}$ into a two-dimensional vector, $G$ is a two-dimensional vector of Gumbel noise \cite{jang2016categorical}, and we abuse the softmax notation to indicate the corresponding probability of being equal to~$1$. We then define $g$ as the estimated total number of processing steps performed, which is calculated as the sum of all the output values $\mathcal{G}_{\phi} \left( \mathbf{v}_{j - 1} \right), \enskip j \in [1, n]$:
\begin{equation}
\label{eq:total_num_iteration}
    \begin{gathered}
    g = {\textstyle \sum}_{j=1}^n \mathcal{G}_{\phi} \left( \mathbf{v}_{j-1} \right)
    \end{gathered}.
\end{equation}
The gating module is trained along with the separation network using an auxiliary loss $\mathcal{L}_{\mathcal{G}}$ based on $g$. This loss is then added to the loss which quantifies separation performance $\mathcal{L} \left(\hat{\s}, \s \right)$. During inference, Gumbel noise is not used in the gating module, meaning that the choice of the gating module is deterministic. Therefore, we note that once a skip decision has been taken (i.e., $\mathcal{G}_{\phi} \left( \mathbf{v}_{i-1} \right) = 0$), we can take an early exit as shown in Fig.~\ref{fig:gating}.

\section{Experimental Setup}
\label{sec:setup}

\subsection{Dataset}
For our experiments, we use the WHAM! dataset \cite{wichern2019wham}, which is constructed using two-speaker utterances from the wsj0-2mix dataset \cite{hershey2016deepclustering} and non-stationary ambient noise sources. The dataset consists of 20,000, 3,000, and 5,000 training, validation, and test samples, respectively. The two speaker utterances are mixed with a uniform random input SNR sampled between the range $\mathcal{U}[0,5]$ dB, while the SNR value between the first (louder) speaker and the noise waveform is similarly sampled from $\mathcal{U}[-3,6]$ dB. 
% We use the min version of WHAM! where the mixtures are truncated to the length of the shorter of the two speech signals. 

We perform experiments on two tasks, namely, noisy speech separation and speech enhancement, where each mixture consists of two utterances plus the additive noise and one speech utterance plus the noise waveform, respectively.
% and speech separation. In  speech enhancement experiments only the first speaker is used in addition with the noise. In contrast, in the separation experiments we use the noisy mixtures with both speakers. 

\subsection{Separation Network}

We choose the \textit{Sudo rm -rf} \cite{tzinis2022compute} architecture for our experiments, as its performance is close to state-of-the-art architectures while requiring lower computational and memory complexity. The encoder and decoder modules use a kernel size of 21 samples and stride of 10 samples, while the number of bases of the encoder and the decoder is set to 512. Each U-ConvBlock serves as a sub-block and consists of 5 successive downsampling and upsampling operations. 

% In the case of speech enhancement we select 2 as the number of output sources with the first being a  dedicated speech source while the second is the noise source. In the separation task we use 3 output sources where the first two are speech outputs, while the third is a dedicated noise output.

%We underline that our proposed scheme is not bounded to this specific architecture and can be effortlessly applied to other architecture with... 

\subsection{Training and Evaluation Details}

We train all models for 200 epochs on 4-second mixture chunks (zero-padded if needed) sampled at 8 kHz. We use the Adam optimizer \cite{adam} with an initial learning rate set to $10^{-3}$ which decays to one-third of its previous value every 40 epochs. Additionally, we perform gradient clipping when the $L_2$-norm exceeds 5. We select a batch size of $B = 4$ and we perform online data augmentation \cite{tzinis2020twostep} within each batch. 
% Specifically, we follow the process introduced in \cite{tzinis2020twostep} where the batch consisting of $B$ groups of sources, has the order of those groups shuffled, while the relative energies between each of the sources at each batch position is kept fixed.
All models are trained to minimize the negative scale-invariant signal to distortion ratio (SI-SDR) \cite{leroux2019sdr} loss, computed between the estimated source $\hat{\s}$ and the clean source $\s$ which is defined as:
\begin{equation}
\label{eq:SISDRloss}
    \begin{gathered}
    \text{SI-SDR}(\hat{\s}, \s) = 10 \log_{10} \left(\| \rho \s\|^2 / \| \rho \s - \hat{\s}\|^2 \right),
    \end{gathered}
\end{equation}
where $\rho =  \hat{\textbf{s}}^\top  \textbf{s} /\|\textbf{s}\|^2$ is a scalar. The loss function used during speech separation training is the permutation invariant \cite{Isik2016Interspeech09,Yu2017PIT} negative SI-SDR, where permutations are only considered on the speech sources. For the enhancement task, we use the mean negative SI-SDR between the estimated speech and the ground-truth target. 

During model evaluation, we report the mean SI-SDR improvement (SI-SDRi) of the speech sources, which quantifies the improvement the estimated sources $\hat{\s}$ achieve over using the input mixture signal as a source estimate.

\section{Results \& Discussion}
\label{sec:results}

\subsection{Iterative Block-Wise Latent Refinement}
In the first set of experiments, we compare the performance and model size of the block-wise iterative method (as presented in Section \ref{subsec:blockiter}) against conventional non-iterative models with various numbers of blocks, sub-blocks, and iterations. 
% Most of the experiments we perform consist of a single block that is iterated a set number of times. 
Each single block can be conveniently defined using several U-ConvBlocks (sub-blocks) from the {Sudo rm -rf} architecture. For example, a conventional Sudo rm -rf network with $k$ U-ConvBlocks (each with unique parameters) can alternatively be viewed as a single block containing $k$ sub-blocks and without iterations. 
% We perform one additional experiment with 2 blocks which contain 4 sub-blocks and are iterated 2 times. 

\begin{table}[!tb]
    \caption{Mean test SI-SDR (dB) performance for block-wise iterative latent refinement for the tasks of speech enhancement (Enh.) and separation (Sep.) where each block consists of several sub-blocks and might be used for a number of iterations. Non-iterative models are denoted with an $\xmark$ in the iterations columns.}
    \label{tab:iter_performance}
    \centering
    \sisetup{
    detect-weight, % Make siunitx detect align bold cells correctly
    mode=text, % Make siuntix print tables in text mode (causes width of bold characters to be the same as non-bold)
    tight-spacing=true,
    round-mode=places,
    round-precision=1,
    table-format=2.1
    }
    \resizebox{\linewidth}{!}{
    \begin{tabular}{ccc*{2}{S}*{2}{c}}
    \toprule
    \multicolumn{3}{c}{Model Configuration} & \multicolumn{2}{c}{SI-SDRi (dB)}& \multicolumn{2}{c}{Params (M)} \\ 
         \cmidrule(lr){1-3}\cmidrule(lr){4-5}\cmidrule(lr){6-7}
    Blocks & Sub-Blocks & Iter. & {Enh.} & {Sep.} & {Enh.} & {Sep.}  \\ 
    \midrule
    \multirow{10}{*}{1}  & \multirow{5}{*}{1} & \xmark & 10.2 & 7.1 & {\multirow{5}{*}{0.4}} & {\multirow{5}{*}{0.5}}  \\
     &  & 2 & 11.5 & 8.8 &  &  \\
     & & 3 &  12.0  &  9.5 &   &   \\
     & & 4 & 12.3 &  10.0 & &     \\
     & & 16 & 13.6  &  12.3 & &    \\ \cmidrule(lr){2-7}
     & \multirow{4}{*}{4} & \xmark & 13.0 & 10.9 & {\multirow{4}{*}{0.8}} & {\multirow{4}{*}{1.0}} \\
     &  & 2 & 13.6 & 12.2  & & \\
     &  & 3 &  14.0 &  12.9 &  & \\
     & & 4 & 14.2 &  13.6 & &   \\ \cmidrule(lr){2-7}
     & 8 & \xmark & 13.8 & 12.6 & 1.4 & 1.5  \\ \cmidrule(lr){2-7}
     & 16 & \xmark & 14.4 & 13.8 & 2.6 & 2.7  \\
    \midrule
    \multirow{1}{*}{2}
     & 4 & 2 & 14.3 & 13.9 & 1.4 & 1.5  \\
    \midrule
    \multicolumn{3}{c}{Sepformer \cite{subakan2021attention}} & 14.4 & 16.3 & 25.7 & 25.7  \\
    \bottomrule
    \end{tabular}
    }
    % \vspace{-5pt}
\end{table}

In Table \ref{tab:iter_performance}, we summarize the results for our iterative block-wise latent refinement method versus conventional non-iterative models under both speech separation and speech enhancement setups. Looking at the results with $1$ block, more iterations lead to significantly higher SI-SDR performance without increasing the trainable parameters of the overall model: $10.2 \rightarrow 13.6$ dB and $7.1 \rightarrow 12.3$~dB for separation and denoising, respectively. Overall, all models that re-use their blocks lead to results similar to those of the conventional models while requiring a much smaller model size, making them more amenable to deployment on edge devices. Moreover, when increasing the representational capability of the block using four sub-blocks, we obtain a separation performance close to the state-of-the-art Sepformer \cite{subakan2021attention}, which contains $30\times$ more parameters and requires significantly more memory and computational resources. The latter result highlights the importance of employing sophisticated iterative latent refinement strategies to better understand and reduce the redundant sub-structure of source separation networks.

\subsection{Progressive Latent Representation Training}
In the next set of experiments, we test the proposed progressive training scheme (as presented in Section \ref{subsec:progtrain}). Our aim is to compare it against end-to-end training in terms of performance and memory requirements during training. An additional goal is to investigate the effect of the number of training steps. We note that the number of progressive training steps coincides with the number of blocks. The blocks are in some experiments iterated and contain a number of sub-blocks which again correspond to U-ConvBlocks. We split the total 200 training epochs evenly among the progressive training steps, and the learning rate schedule is reset at the start of each step. 
\begin{table}[!htb]
\caption{Progressive Training Results. The number of blocks corresponds to the number of progressive training steps and each of those blocks contains a set number of sub-blocks. The number of iterations refers to the number of times each block is iterated. The right-most columns indicate the maximum memory required in GBs during a backward pass with a batch size of 1.}
    \centering
    \sisetup{
    detect-weight, % Make siunitx detect align bold cells correctly
    mode=text, % Make siuntix print tables in text mode (causes width of bold characters to be the same as non-bold)
    tight-spacing=true,
    round-mode=places,
    round-precision=1,
    table-format=2.1
    }
    \resizebox{\linewidth}{!}{
    \begin{tabular}{ccc*{2}{S}*{2}{S[round-precision=2,table-format=1.2]}}
    \toprule
     \multicolumn{3}{c}{Model Configuration} & \multicolumn{2}{c}{SI-SDRi (dB)} & \multicolumn{2}{c}{GPU Mem (GB)} \\ 
     \cmidrule(lr){1-3}\cmidrule(lr){4-5}\cmidrule(lr){6-7}
    Blocks & Sub-Blocks & Iter. & ~{Enh.} & {Sep.} & ~~{Enh.} & {Sep.}  \\ 
    \midrule
    %\cline{1-7}
    \multirow{3}{*}{2} & 8 & \xmark & 14.0 & 13.8 & 1.06 & 1.09  \\
    & 4 & 2 & 14.0 & 13.5 & 1.05 & 1.08 \\
    & 2 & 4 &  13.8 &  13.3 & 1.05 & 1.07 \\ 
    \midrule
    %\cline{1-7}
    \multirow{3}{*}{4} & 4 & \xmark & 13.6  &  12.9 & 0.58 & 0.61 \\ 
    & 2 & 2 & 13.4 & 12.7 & 0.58 & 0.60 \\
    & 1 & 4 & 13.2 & 12.2 & 0.57 & 0.60\\ 
    \midrule
    %\cline{1-7}
    1 & 16 & \xmark & 14.4 & 13.8 & 2.02 & 2.05 \\
    \bottomrule
    \end{tabular}
    }
    \label{tab:progressive_performance}
    % \vspace{-5pt}
\end{table}

We show the separation performance and the actual memory footprint of the proposed progressive training versus the conventional end-to-end training in Table \ref{tab:progressive_performance}. Progressive training leads to marked improvements in terms of lowering the overall training footprint of separation models which unequivocally enables much higher data-processing parallelization on the training device by allowing the usage of much larger batch sizes. Specifically, when progressively training two blocks with the application of various iterations of each block, we are able to obtain similar separation and denoising performance while halving the total training memory footprint. By increasing the number of progressively trainable blocks to $4$ but decreasing the total amount of sub-blocks and iterations in order to keep a constant inference FLOP count, we obtain a remarkable decrease in the overall required training memory bottleneck while observing only a slight performance hit in terms of absolute SI-SDR. 

\subsection{Adaptive Early Exit}
% In the last experiment, we propose the use of a gating module at the input of each block to enable adaptive inference by dynamically controlling the number of block-wise iterations performed given an input noisy speech mixture for the task of speech enhancement. 

In the last experiment, we propose to use a gating module at the input of each block which dynamically adapts the computation performed for each input mixture. In essence, this adaptive inference scheme decides the number of processing steps performed.
%blocks used and  how many times each one is going to be iteratively applied. 

% enable adaptive inference by dynamically controlling the number of block-wise iterations performed given an input noisy speech mixture for the task of speech enhancement. 
% We attach a gating module to the model consisting of 1 block containing 1 U-ConvBlock iterated a maximum of 4 times.

The gating module first transforms the latent representation $\mathbf{v}_i$ of size ${128 \times 3200}$ into a two-dimensional vector using the function $f$, and then into a binary decision according to Eq. \ref{eq:gating}. For the function $f$, we choose a sequence of a Conv1d layer with a kernel of size 1 with 128 input and 2 output channels, a PReLU activation and another Conv1d layer with kernel size of 3200 and 2 input and 2 output channels. The total number of parameters of the gating module is 13K, whereas a single U-ConvBlock contains 147K parameters. The FLOPS overhead is also negligible, namely, a 0.2\% increase. 

The network that we attach the gating module onto consists of one block containing one U-ConvBlock iterated up to 4 times. This corresponds to the description of Section \ref{subsec:adaptive} with all blocks having shared parameters. The network without the gating module is pre-trained for 180 epochs. Then both the network and the gating module are jointly trained for 20 more epochs. The initial learning rate is set to $10^{-4}$ and decays every five epochs to one-third of its previous value. Finally, the overall loss function penalizes bad reconstruction quality as well as enforces the model to use less iterations:
\begin{equation}
\label{eq:loss_early_exit}
    \begin{aligned}
    \mathcal{L} = \mathcal{L}_{\text{SI-SDR}} + \mathcal{L}_{\mathcal{G}}, \enskip \mathcal{L}_{\mathcal{G}} = 0.75 \left(g - 3 \right)^2,
    \end{aligned}
\end{equation}
where $g$ is the estimated number of iterations performed (see Eq.~\ref{eq:total_num_iteration}). % calculated according to Eq.~\ref{eq:total_num_iteration}.

\begin{figure}[!h]
    \centering
    \includegraphics[trim={0cm 0 0cm 0},clip,width=\linewidth]{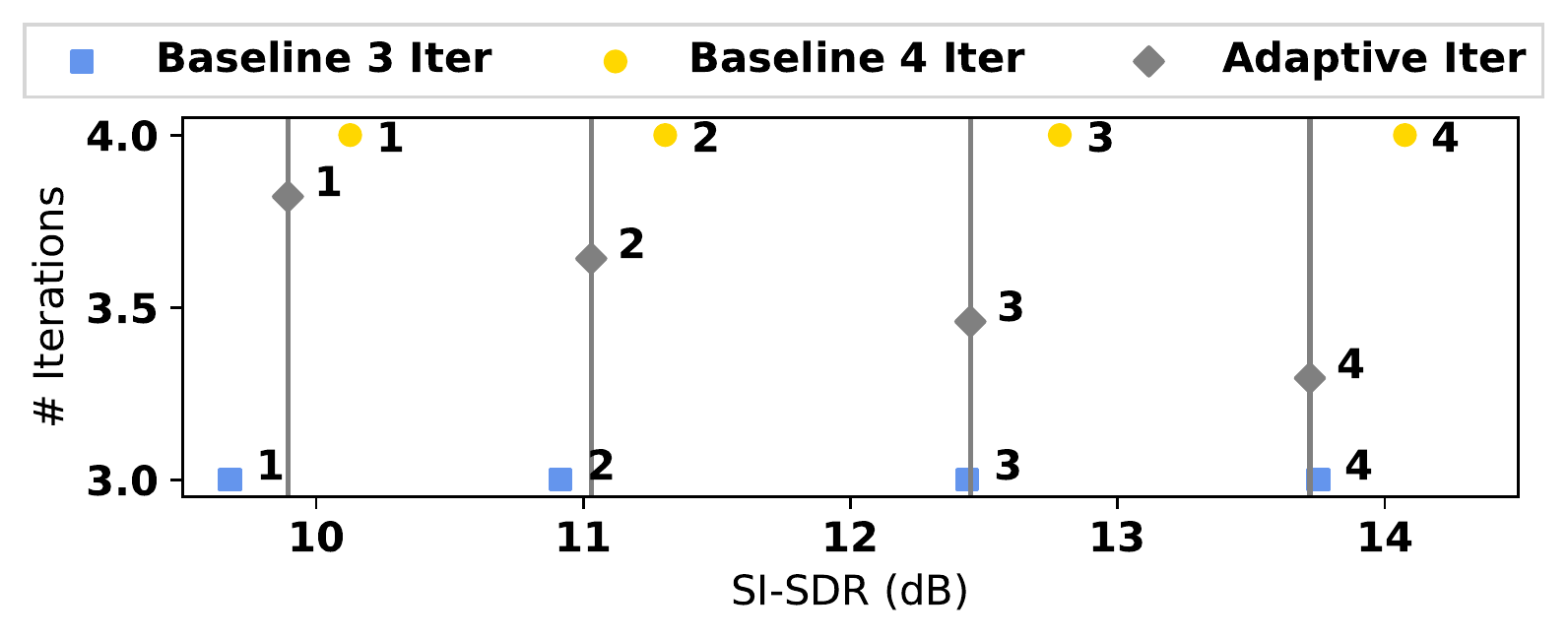}
      \caption{Comparison of the adaptive model (grey) against the baseline models (blue and gold) that have been trained to perform 3 and 4 iterations respectively.  Scatter plot of mean SI-SDR performance and mean number of iterations of samples grouped by their SNR. Points are annotated with the corresponding quartile bin. For example, points annotated with 1 represent a group of samples with SNR in the bottom 25\%, while points annotated with 4 represent the top 25\%. Each marker color/shape represents a different model. }
      \label{fig:earlyexitres}
    % \vspace{-8.8pt}
\end{figure}  

In Fig.~\ref{fig:earlyexitres}, we show the behavior of the model with the adaptive gating module.
%, which consists of one block containing one U-ConvBlock iterated up to 4 times. 
We observe that the model manages to distinguish samples based on their input SNR and adapt to them, while performing 3.6 iterations on average. Specifically, the gating module chooses to perform on average more (less) iterations for samples with lower (higher) input SNR, thus, saving computational resources and potentially being able to better fit the data. Moreover, we compare the adaptive model against two baseline models with a) 1 block with 3 iterations and b) 1 block with 4 iterations. First, we note that in terms of mean SI-SDR and in most of the quartile bins, the adaptive model outperforms the baseline with 3 iterations. Compared to the baseline with 4 iterations, the adaptive model leads to a small separation performance drop but this can be explained due to the fact that the adaptive model has to maintain high-level performance for both 3 and 4 iterations, while the baseline model is allowed to focus only on performing well for 4 iterations.

% \begin{table}[!htb]
%     \centering
%     \begin{tabular}{ccc|cc|cc}
%     \toprule
%     \multicolumn{3}{c|}{Model Config} & \multicolumn{2}{c|}{Mean Test SI-SDRi (dB)}& \multicolumn{2}{c}{Model Size (M)} \\ \cline{1-7}
%     SB & B & Iter & Enh & Sep & Enh & Sep  \\ \cline{1-7}
%     -- & 8 & --- & $13.8$ & $12.6$ & $1.4$ & $1.5$  \\ \cline{1-7}
%     2 & 4 & 2 & $14.3$ & $13.9$ & $1.4$ & $1.5$  \\ \cline{1-7}
%     -- & 16 & --- & $14.4$ & $13.8$ & $2.6$ & $2.7$  \\
%     \bottomrule
%     \end{tabular}
%     \caption{Block Iterations}
%     \label{tab:iter_performance2}
%     % \vspace{-5pt}
% \end{table}

\section{Conclusion}
\label{sec:conclusion}
We have provided a general formulation for source separation model training as a latent variable refinement process. Our holistic approach enables several optimizations and improvements over conventional models and end-to-end training by explicitly leveraging the repetitive structure of processing blocks inside a neural network. Our experiments showcase the effectiveness of our approach in terms of providing a) a much lighter computational model by leveraging the application of processing blocks multiple times, b) a much lower training memory footprint by using progressive training, and c) a much more flexible inference graph which can dynamically adapt its structure given the input data. %In the future, we aim to provide more efficient latent variable refinement strategies based on meta- and multi-task learning methods. 

\balance
\bibliographystyle{IEEEtran}
\bibliography{refs}

\end{document}